# Homogeneous Nucleation and Forest Hardening Result in Thermal Hardening Phenomenon in Shock Loaded BCC Metals

Songlin Yao, Xiaoyang Pei, Jidong Yu[1], Yuying Yu, Qiang Wu[2]

***Abstract***

Interpretation of "thermal hardening" phenomenon at high strain rate has recently become a critical problem in shock wave physics. In this letter, this problem is addressed from a viewpoint of dislocation generation, and a novel conclusion is gained that forest hardening induced by homogeneous nucleation (HN) results in thermal hardening behavior in a BCC metal significantly, apart from phonon drag mechanism. Through numerical simulations with a dislocation based crystal plasticity model, we have reproduced the experimental results quantitatively and predicted a thermal hardening behavior in other BCC metals, i.e., Mo, at higher temperature.

***Introduction.-***

Strength of a metal is usually regarded as thermal softening since the strength vanishes above the melting temperature[12]. However, thermal hardening phenomenon has been widely observed in shock loading experiments, and has recently become a critical problem in shock wave physics[3]. It was observed experimentally in shock loaded FCC and HCP metals that the typical features of the velocity profiles, including both the Hugoniot Elastic Limit (HEL) spike and the HEL minimum, present thermal hardening phenomenon[4567]. The HEL, also known as dynamic strength[8], marks the onset of shock induced plasticity. Regarding to BCC metals, most BCC metals still present thermal softening behaviors[9101112] except V and the ferromagnetic metal Fe. In particular, the HEL spikes of V present temperature insensitivity, while the HEL minimums present thermal hardening at elevated temperature.

In conventional conceptualization, the extraneous strength stems from dislocation motion. For example, thermal softening behaviors observed at moderate strain rates are attributed to thermally activated dislocation motion[13141516171819], while thermal hardening behaviors of FCC metals subjected to high strain rates are attributed to phonon drag[20]. As opposed to dislocation motion, dislocation generation has also been regarded as the main contributing factor of the strength as strain rate increases. In particular, F-R sources and cross-slip are common dislocation generation mechanisms at moderate strain rates[21], while homogeneous nucleation has long been proposed as a likely mechanism responsible for the generation of dislocations at high strain rates[22232425]. Obviously, thermally activated homogeneous nucleation will predict a thermal softening behavior, like AAZ model did[26]. When dealing with the temperature dependent behaviors of BCC metals under shock loading, Gurrutxaga-Lerma pointed out that the temperature dependent behavior of BCC Fe is a result of the interaction between dislocation generation and dislocation motion, including phonon drag hardening, Peierls stress and enhanced dislocation density.[27]

Besides above theories, a novel explanation of the thermal hardening phenomenon is recently proposed that the main contributing factor to thermal hardening behavior of FCC metals is due to shear modulus modulating the magnitude of the shielding fields emanated from moving dislocations. Nonetheless, this theory seems not suitable for BCC V. The shear modulus of V shows a singular behavior between 800K and 1100K[28], which makes the shear modulus effect invalid.

Above all, we are still confused with the thermal hardening behavior of BCC V. Is phonon drag responsible for the thermal hardening behavior in BCC V? In this letter, we show that forest hardening induced by homogeneous nucleation (HN) results in the thermal hardening behaviors in a BCC metal significantly, while thermal softening behaviors are mainly due to the temperature effect of Peierls stress through a dislocation based crystal plasticity constitutive model. This thermal hardening mechanism proposed in this letter quite differs from existing theories, and has also been proved to be suitable for other BCC metals.

***Classical opinions.-***

A dislocation based crystal plasticity constitutive model is applied to address this problem. (See the supplemental material) In this model, dislocation mobility is controlled by Peierls stress and phonon drag, and dislocation generation is controlled by homogeneous nucleation and multiplication.

[1] Corresponding Author E-mail: yujidong@caep.cn(J. Yu);
[2] wuqiang@caep.cn(Q. Wu)

In classical picture, Peierls stress marks the transition from thermally activated glide of dislocations below Peierls stress and phonon drag mechanism above Peierls stress[29]. When dealing with a weak shock loading problem, the applied stress induced by shock compression exceeds the Peierls stress quickly, leading to the thermally activated glide of dislocations invalid and phonon drag dominating dislocation motion. Thus the dislocation velocity is mainly controlled by the Peierls stress and the phonon drag effect. As temperature increases, drop of Peierls stress makes it easier for a dislocation to move, while growth of phonon drag makes it harder. It seems to us that the interaction between Peierls stress and phonon drag may indeed result in a transition from thermal softening to thermal hardening.

Firstly, let's neglect the temperature dependent dislocation generation, i.e., HN, to check out whether temperature dependent dislocation mobility is responsible for the temperature dependent behaviors of the dynamic yield strength in BCC metals. As shown in FIG. 1, calculated results of both metals at room temperature are in good agreements with experimental results, while calculated HEL minimum and the plastic front of V at elevated temperature deviate from the experimental result obviously. In particular, calculated HEL minimum at 1100K is much smaller than experimental result. Apparently, a constitutive model, in which only the temperature effect of dislocation mobility is taken into account, is incapable of capturing the essence of the plastic deformation in a shock loaded BCC metal at elevated temperature correctly.

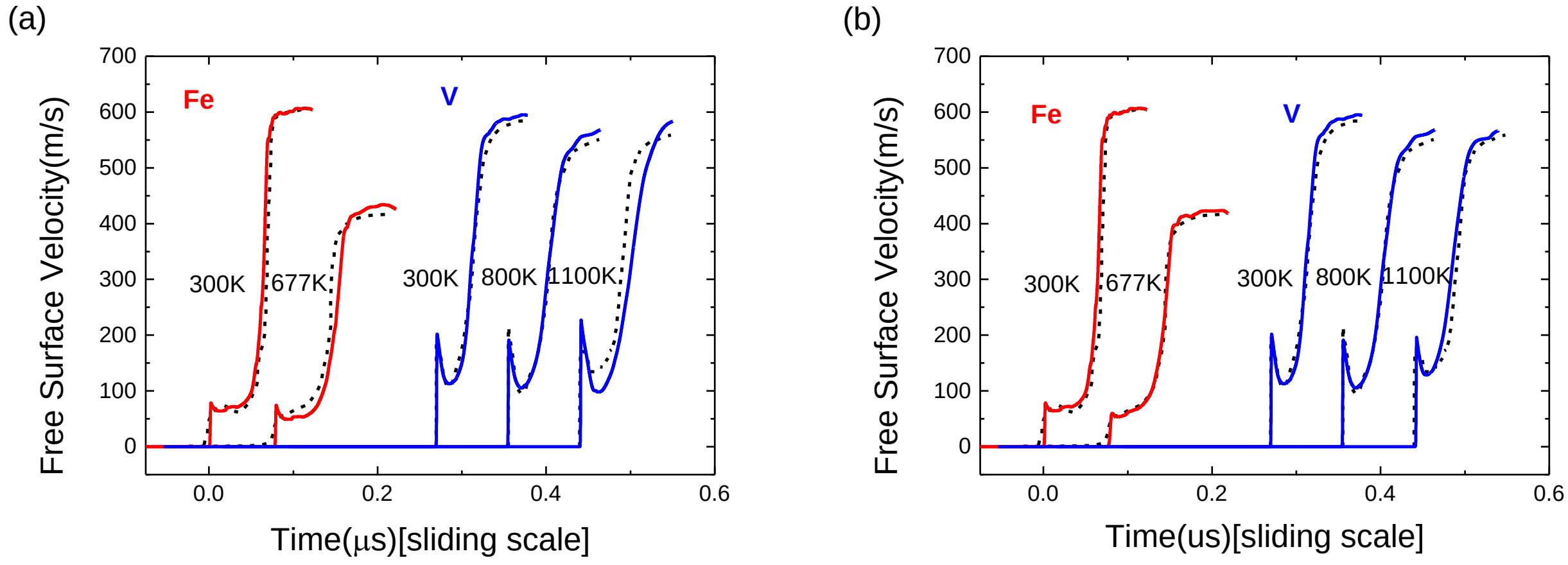

Figure 1 Comparison between experimental results and calculated results: a) without HN; b) with HN. Dashed lines refer to experimental results, and solid lines refer to simulations

***HN.-***

In this letter, temperature dependent dislocation generation is attributed to homogeneous nucleation. HN occurs when applied stress approaches the nucleation strength. The nucleation strength of a metal is about G/15[30], and hard to be achieved[31]. Thus HN is usually thermally activated. Due to its thermally activated property, HN is of high temperature and high stress sensitivity. As temperature increases, decrease of the activation energy and increase of the thermal fluctuation make HN to occur more easily. As shown in FIG. 2, nucleation probability at a single nucleation site grows sharply from almost negligible at room temperature to about 0.2 at 1100K. As the shear stress increases to 2GPa, the probability of nucleation reaches almost 0.4.

Subsequently, we calculated the velocity profiles using a model with thermally activated HN considered, as shown in FIG. 1(b). Calculated results with HN considered show a higher HEL minimum and match better with experimental results, which demonstrates that thermally activated HN is responsible for the thermal hardening behaviors of BCC metals at elevated temperature. Nonetheless, from the viewpoint of dislocation generation, HN is still a thermal softening mechanism, but not a thermal hardening mechanism. How does HN result in the increase of the HEL minimum at elevated temperature?

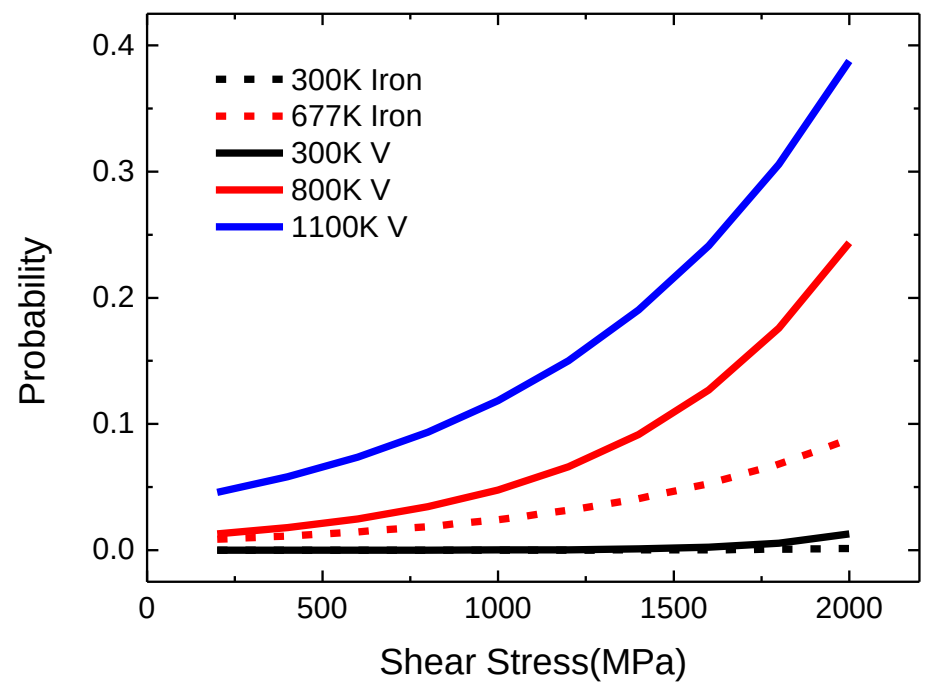

Figure 2 HN probability at a single nucleation site, predicted by equationX

***Forest hardening.-***

Dislocation generation not only contributes to plastic dissipation, but also contributes to hardening via dislocation-dislocation interaction, also known as forest hardening[3233]. Forest hardening and Peierls stress serve as the mechanical threshold stress, also known as the critical resolved shear stress (CRSS), for a dislocation to move together.[34] The dislocation-dislocation interaction serves as the main resistance for a dislocation to move in a BCC metal subjected to high temperature. Existing studies indicate that the plastic hardening of most BCC metals is hardly influenced by temperature, and contributes to the athermal part of the flow stress[35]. However, massive dislocation generation, due to homogeneous nucleation, at high strain rate changes this circumstance. We can learn from time histories of RSS and CRSS that after shock wave arrives, the CRSS of both metals at room temperature almost doesn't evolve, while the CRSS at elevated temperature increases considerably. As shown in FIG. 3(b), the CRSS of V at 1100K even exceeds the CRSS at 300K at the HEL minimum. Since it's assumed that the Peierls stress is determined by the ambient temperature, increase of the CRSS is then due to the forest hardening. Massive dislocations are generated at elevated temperatures due to strong temperature sensitivity of HN[36], leading to thermal hardening of the dislocation-dislocation interaction. That's how HN controls the thermal hardening of the HEL minimum of V. In this case, why was not the same phenomenon observed in other shock loaded BCC metals?

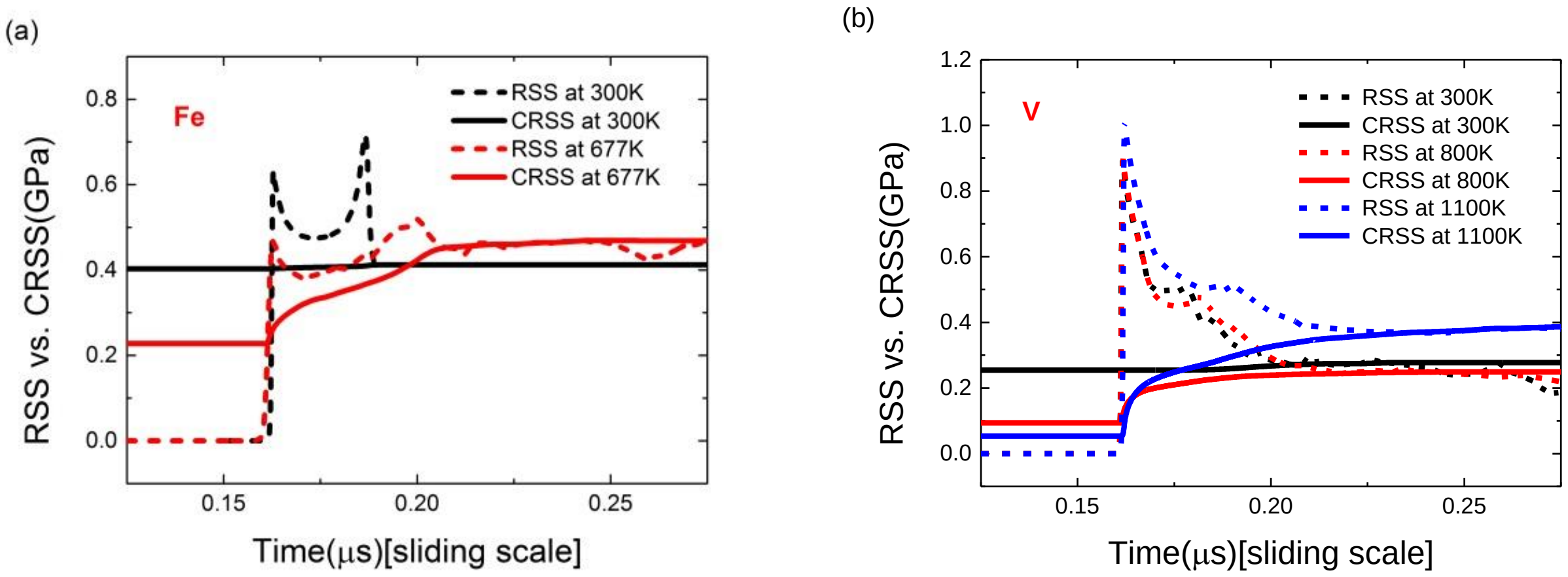

Figure 3 Time histories of RSS and CRSS: a) Fe; b) V

***Forest hardening and Peierls stress.-***

Though forest hardening plays the same role as the Peierls stress does on plastic deformation, their temperature effects on dislocation mobility happen to be opposite to each other. As temperature increases, rapid decrease of Peierls stress contributes to thermal softening of the dynamic yield strength, while increase of the foresting hardening contributes to thermal hardening of that.

Based on the Kuhlmann-Wilsdorf theory[37], Peierls stress of most BCC metals drop sharply with temperature. Thermal hardening of the forest hardening is not enough to make up for the loss of the Peierls stress as temperature increases. Thus Peierls stress dominates the temperature dependent behavior of CRSS, leading to a thermal softening behavior. That's why Fe and V present thermal softening behavior below 677K and below 800K respectively. As temperature increases, Peierls stress almost vanishes, as shown in FIG. 4. Forest hardening resistance, induced by HN, dominates the temperature dependent behavior of CRSS, leading to a thermal hardening behavior. As shown in FIG. 3, the CRSS of V at 1100K grows sharply and exceeds the CRSS at 300K after shock wave arrives. Moreover, phonon drag also contributes to the thermal hardening behavior. Consequently, V presents thermal hardening behavior above 800K.

It seems to us that this thermal hardening mechanism is also suitable for FCC metals. Using this model, we simulated the temperature dependent response of weak shock wave loaded FCC metals. However, the simulated results indicate that this mechanism doesn't influence the profile evolution significantly (see the supplemental materials). We attributed to this difference between FCC and BCC metals to different dislocation mobility. Lower viscosity of FCC metals makes it easier for a dislocation to slip, and higher dislocation velocity. More energy is dissipated via dislocation slip, while less is dissipated via dislocation nucleation. Thus HN is not significant, neither does the forest hardening.

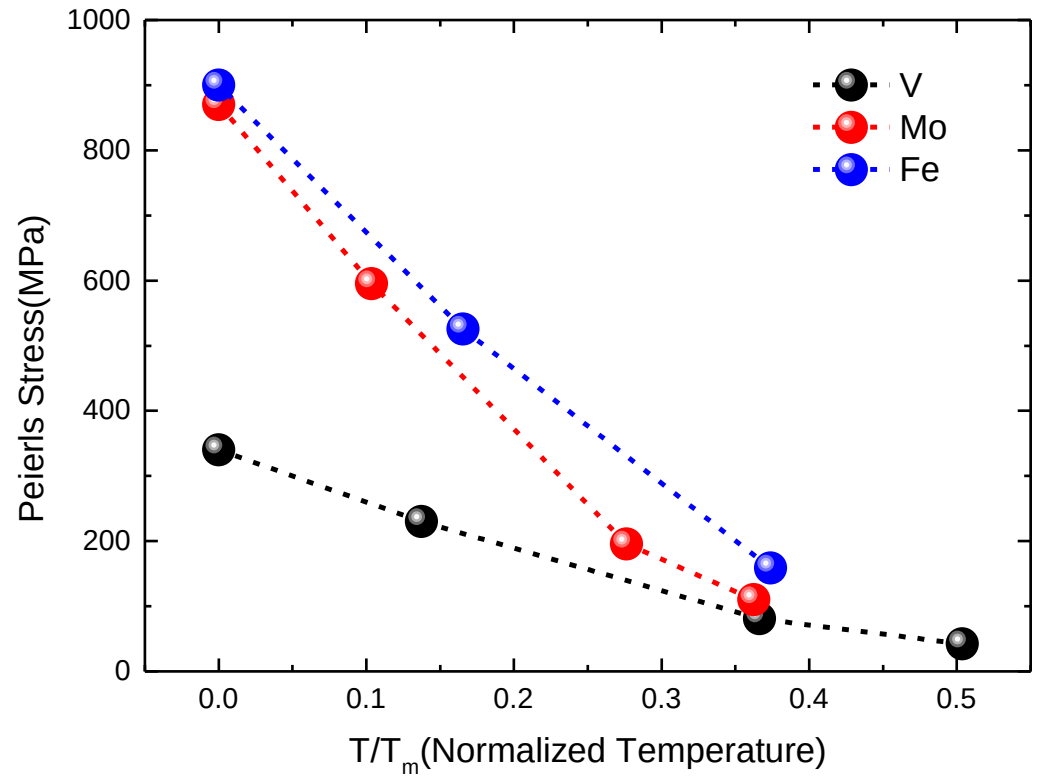

Figure 4 Peierls stress vs. temperature

***Prediction.-***

Based on above analysis, it's expected that the thermal hardening phenomenon could also be observed in other BCC metals at higher temperature, at which the Peierls stress vanishes and forest hardening resistance dominates the CRSS. To do such a simulation, the saturation dislocation density[38] is taken into account in case that the dislocation density generated by HN at elevated temperature exceeds the saturation density quickly. The saturation density indicates that the temperature effect of HN is not significant any more as the temperature exceeds a threshold. The only left mechanism is phonon drag mechanism, which predicts a thermal hardening behavior. Calculated results are displayed in FIG.5, and are in accordance with experimental results and theoretical analysis. The same behavior is expected to be suitable for other BCC metals, phonon drag effect of which is temperature sensitive at elevated temperature.

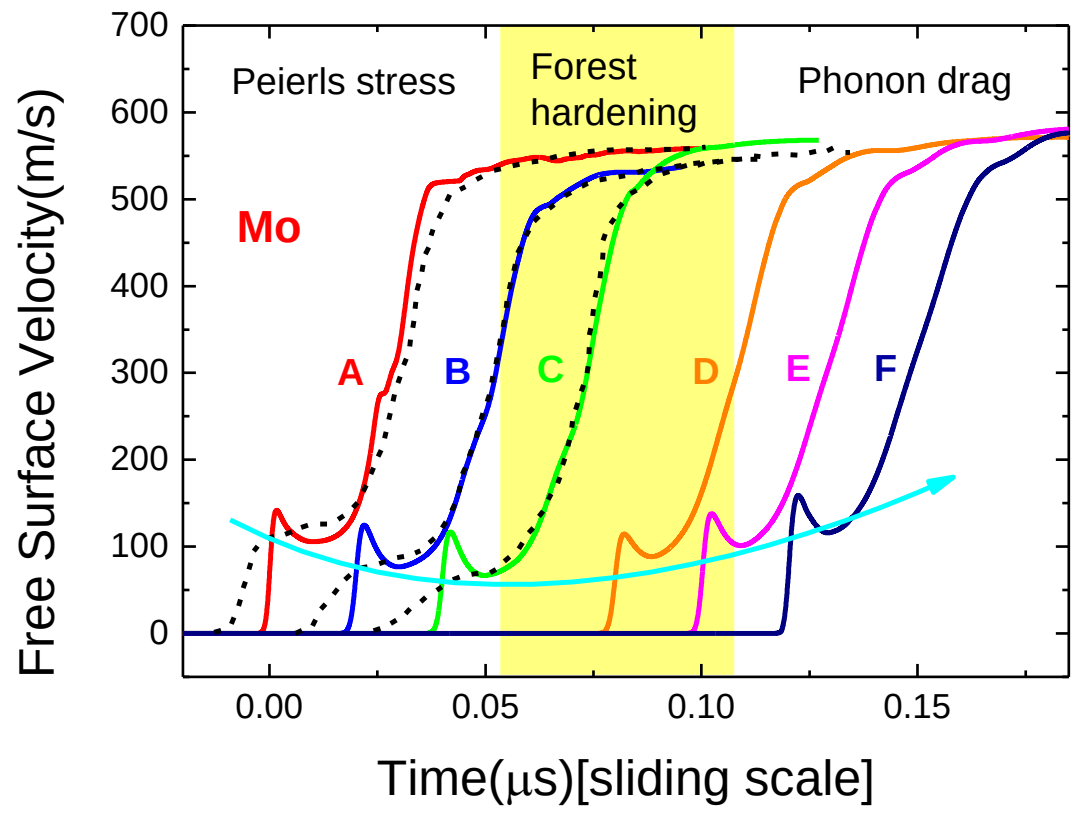

Figure 5 Dashed lines refer to experimental results, and solid lines refer to simulated results and predictions, A, B, C, D, E, F refer to simulated results at 300K, 800K, 1050K, 1500K, 2000K, and 2500K respectively

Moreover, the dominant mechanism of dislocation evolution in BCC metals could be sorted into three regimes according to temperature, i.e., Peierls stress controlling mechanism, forest hardening controlling mechanism and phonon drag controlling mechanism. At lower temperature, quick drop of Peierls stress dominates the temperature effect of CRSS, leading to a thermal softening phenomenon of the strength. As temperature increases, the Peierls stress almost vanishes, forest hardening induced by HN contributes to thermal hardening of the strength. At even higher temperature, dislocation density approaches the saturation density quickly due to HN, resulting in the temperature effect of HN on forest hardening invalid. Phonon drag results in a thermal hardening behavior.

***Summary.-***

We have shed light on the underlying mechanism that results in the thermal hardening phenomenon in shocked BCC metals through a dislocation based model. The novelty of this work is that the temperature dependent behaviors of the strength at high strain rates are addressed from a viewpoint of dislocation generation. It's concluded that forest hardening induced by HN contributes to thermal hardening of the strength of BCC metals significantly, apart from the classical theory that phonon drag leads to the thermal hardening phenomenon solely. We believe that this study is profitable to the interpretation of experimental data at extreme conditions and the refinement of the theory of dynamic deformation.

***Acknowledgements.-***S. L. Yao has benefited a lot from discussions with R. W. Armstrong and J. Wang. The authors would like to thank W. Q. Wang, P. Li, H. Y. Geng, Z. L. Liu, L. Y. Kong and K. G. Chen for useful remarks, and S. Chen for help of language. This work was supported by Science Challenge Project (Grant No. TZ2018001), the Foundation of the President of the China Academy of Engineering Physics (Grant No. 201402084), and the National Natural Science Foundations of China (Grant Nos. 11302202 and 11532012).

[1] S. Nemat-Nasser, and Y. Li, Acta Mater. **46**, 2(1998)
[2] G. Z. Voyiadjis and A. H. Almasri, Mech. Mater., **40**, 6(2008)
[3] G. I. Kanel, J. Phys. Conf. Series, **500**, 1(2014)
[4] G. I. Kanel et al., J. Appl. Phys., **90**, 1(2001)
[5] E. B. Zaretsky and G. I. Kanel, J. Appl. Phys., **110**, 7(2011)
[6] E. B. Zaretsky and G. I. Kanel, J. Appl. Phys., **114**, 8(2013)
[7] G. I. Kanel et al., **116**, 14(2014)
[8] M. A. Shehadeh and H. M. Zbib, Philosophical Magazine, **96**, 26(2016)
[9] E. B. Zaretsky and G. I. Kanel, J. Appl. Phys., **115**, 24(2014)
[10] E. B. Zaretsky and G. I. Kanel, J. Appl. Phys., **117**, 19(2015)
[11] E. B. Zaretsky and G. I. Kanel, J. Appl. Phys., **120**, 10(2016)
[12] E. B. Zaretsky and G. I. Kanel, J. Appl. Phys., **122**, 11(2017)
[13] J. Weertman and P. S. Follansbee, Mech. Mater., **7**, 3(1988)
[14] E. Zaretsky, J. Appl. Phys., **78**, 6(1995)
[15] E. Hornbogen, Acta Metall., **10**, 10(1962)
[16] G. Z. Voyiadjis and F. H. Abed, Mech. Mater., **37**, 2-3(2005)
[17] F. J. Zerilli and R. W. Armstrong, J. Appl. Phys. **61**, 5(1987)
[18] F. J. Zerilli and R. W. Armstrong, J. Appl. Phys. **68**, 4(1990)
[19] M. A. Meyers et al., Mater. Sci. Eng. A, **322**, 1-2(2002)
[20] V. S. Krasnikov, A. E. Mayer and A. P. Yalovets, Int. J. Plast., **27**, 8(2011)
[21] B. Gurrutxaga-Lerma et al., J. Appl. Mech., **82**, 7(2015)
[22] M. Meyers, Scr. Metall., **12**, 1 (1978)
[23] C. Smith, Trans. Met. Soc. AIME, **212**, 10(1958)
[24] M. A. Shehadeh et al., Appl. Phys. Lett., **89**, 17(2006)
[25] B. Gurrutxaga-Lerma et al., J. Mech. Phys. Solids, **84**, 1(2015)
[26] R. W. Armstrong, W. Arnold and F. J. Zerilli, J. Appl. Phys., **105**, 2(2009)
[27] B. Gurrutxaga-Lerma et al., Int. J. Plast., **96**, 135(2017)
[28] E. Walker, Solid State Communications, 28, 1978, 587-589
[29] J. Marian, W. Cai and V. V. Bulatov, Nat. Mater., 3, 3(2004)
[30] J. P. Hirth and J. Lothe, *Theory of Dislocations*, (Academic, New York, 1982), Chap. 20, p. 757-759
[31] M. A. Shehadeh and H. M. Zbib, Philos. Mag., **96**, 26(2016)
[32] M. Tang, M. Fivel and L. P. Kubin, Mater. Sci. Eng. A, **309-310**, 256(2001)
[33] S. Queyreau, G. Monnet and B. Devincre, Int. J. Plast., **25**, 2(2009)
[34] R. Madec, B. Devincre and L. P. Kubin, Phys. Rev. Lett., **89**, 25(2002)
[35] G. Z. Voyiadjis and F. H. Abed, Mech. Mater., **37**, 2-3(2005)
[36] S. Ryu, K. kang and W. Cai, J. Mater. Res., **26**, 18(2011)
[37] B. I. Smirnov, Mech. Mater., **3**, 4(1967)
[38] N. R. Barton et al., J. Appl. Phys., **109**, 7(2011)